\documentclass[draft,showkeys,showpacs,eqsecnum,nofootinbib]{revtex4}
\renewcommand{\theequation}{\arabic{equation}}
\def\beq{\begin{equation}}
\def\eeq{\end{equation}}
\def\bea{\begin{eqnarray}}
\def\eea{\end{eqnarray}}\def\nn{\nonumber}

\def\na{\nabla}
\def\pa{\partial}

\def\nn{\nonumber}

\begin{document}
\title{Geometrical and hydrodynamic aspects of five-dimensional Schwarzschild black hole}
\author{Soon-Tae Hong}
\email{soonhong@ewha.ac.kr}
\affiliation{Department of Science
Education and Research Institute for Basic Sciences, Ewha Womans
University, Seoul 120-750, Republic of Korea}
\date{\today}
\begin{abstract}
Exploiting the five-dimensional Schwarzschild black hole, we study the geometrical natures of the higher dimensional 
black hole to yield the (6+1) dimensional global embedding Minkowski space structure. We next obtain the Hawking 
temperature on this five-dimensional manifold, whose result is different from the four-dimensional one.
On the contrary, the radial component of the Einstein equation for the massive particles or photons
on the five-dimensional spacetime is shown to have the same form as the four-dimensional black hole one.
Moreover, we construct the effective potential on equatorial plane of the restricted three-brane to investigate the behavior 
of the particles or photons on this restricted brane.
\end{abstract}
\maketitle

\section{Introduction}
\setcounter{equation}{0}
\renewcommand{\theequation}{\arabic{section}.\arabic{equation}}

There have been tremendous progresses in lower dimensional black holes
associated with the ten-dimensional string theory~\cite{green87,pol99} since an exact conformal field
theory describing a black hole in two-dimensional space-time was
proposed~\cite{witten91}. Even though the (2+1) dimensional Ba\~nados-Teitelboim-Zanelli
(BTZ) black hole~\cite{btz1,btz12,btz2,cal,mann93} is a toy model
in some respect, the BTZ black hole has triggered significant
interests due to its connections with some string
theories~\cite{high,high2} on ten-dimensional space-time, its role in microscopic entropy
derivations~\cite{cal95,cal952} and quantum corrected
thermodynamics~\cite{man97,man972,man973}. Specifically, a slightly 
modified solution of the BTZ black hole yields an solution to the 
string theory, so-called the black string~\cite{horowitz93,horowitz932}. 
Here one notes that this black string solution is in fact only a solution to the lowest
order $\beta$-function equation to receive quantum
corrections~\cite{callan85}. Recently, the Hawking temperatures~\cite{hawk75,hawk752} and Unruh
effects~\cite{unr} of the (2+1) dimensional BTZ balck holes have been
analyzed~\cite{hongbtz00,hongst00} and the (2+1) dimensional black strings have been later
investigated~\cite{hong01prd} in the framework of the global embedding Minkowski 
space (GEMS) structures.  

On the other hand, the string singularities, which are the string
theory version of the Hawking-Penrose
singularities~\cite{hawking}, have been applied to the early
universe with an arbitrary higher
dimensionality~\cite{hong07,hong08,hong11}. In this higher
dimensional stringy cosmology, the expansion of the universe has
been explained by exploiting Hawking-Penrose type singularity in
geodesic surface congruences for the timelike and null
strings~\cite{hong07,hong08}. Next, the twist and shear have been
studied in terms of the expansion of the universe. Moreover, as
the early universe evolves with expansion rate, the twist of the
stringy congruence decreases exponentially and the initial twist
value should be large enough to sustain the rotations of the
ensuing universe, while the effects of the shear are negligible to
produce the isotropic and homogeneous universe~\cite{hong11}. By
exploiting the phantom field, the evolution of cosmomogy has been
also studied in higher dimensional spacetime~\cite{hong08c}.

Recently, the warp factor has been applied to higher dimensional theories such as the Randall-Sundrum
model~\cite{randall,randall2,rubakov,ito,ito2} in five-dimensions. The hierarchy problem on the 
four-brane and five-brane has been studied by 
attaching a circle and a sphere to the standard three-brane, 
respectively~\cite{hong13}. The effective masses in their excited spectra on 
the four- and five-branes has been shown to indicate interesting characteristics 
associated with the quantization of the compact circular 
and spherical manifolds, while their lightest effective masses are shown 
to suppress exponentially with respective to the Planck mass, 
similar to the standard three-brane case.

In this paper, keeping in mind the gravity and/or cosmology theories related to the ten-dimensional 
string and in the five-dimensional RS model mentioned above, 
we will study the higher dimensional
general relativity theory by considering the specific five-dimensional Schwarzschild black
hole~\cite{tang63,emparan08}.

This paper is organized as follows. In Section II we introduce the
five-dimensional Schwarzschild black hole metric to study 
the GEMS natures of the five-dimensional Schwarzschild black hole, 
and in Section III we investigate the hydrodynamics of the massive 
particles and photons. In Section IV, we will study an effective potential characteristics on 
equatorial plane of the restricted three-brane. Section V includes the summaries and discussions.
In Appendix A, we study the geometrical properties of
five-dimensional Schwarzschild black hole such as the Christoffel
symbols associated with the five-acceleration. We also list the
Riemann tensors of the five-dimensional Schwarzschild black hole,
which will be used in the Einstein equation of interest.

\section{Global flat embeddings and thermodynamics}
\setcounter{equation}{0}
\renewcommand{\theequation}{\arabic{section}.\arabic{equation}}

After Unruh's work~\cite{unr}, it has been known that a
thermal Hawking effect on a curved manifold~\cite{hawk75} can be
looked at as an Unruh effect in a higher flat dimensional space
time. According to the GEMS
approach~\cite{kasner,fro,goenner,ros}, several
authors~\cite{deser97,deser972,deser99,bec,gon,gib99,hongbtz00,hongst00}
recently have shown that this approach could yield a unified
derivation of temperature for various curved manifolds such as
rotating Ba\~nados-Teitelboim-Zanelli (BTZ)~\cite{btz1,btz12,btz2,cal,mann93} 
for instance. 

In order to investigate the global flat embedding structure of the
five-dimensional Schwarzschild black hole defined on the total
manifold $S^{3}\times {\mathbf R}^{2}$, we consider the
five-metric of the form \beq
ds_{5}^{2}=-N^{2}dt^{2}+N^{-2}dr^{2}+r^{2}d\Omega_{3}^{2},
\label{metric5} \eeq where \bea
N^{2}&=&1-\frac{\mu_{5}}{r^{2}},\nn\\
d\Omega_{3}^{2}&=&d\alpha^{2}+\sin^{2}\alpha
(d\theta^{2}+\sin^{2}\theta d\phi^{2}), \label{n2} \eea and
$\Omega_{3}$ is the solid angle in the three-dimensional compact
sphere $S^{3}$ whose value is $2\pi^{2}$. Here we have three
angles of the three-sphere whose ranges are defined by $0\le
\alpha\le \pi$, $0\le \theta\le \pi$ and $0\le \phi\le 2\pi$. In
the lapse function $N$ we have the five-dimensional Schwarzschild
radius $\mu_{5}$ defined as\footnote{The $d$-dimensional
Schwarzschild radius $\mu_{d}$~\cite{emparan08} is given by
$\mu_{d}=\frac{16GM}{(d-2)\Omega_{d-2}}$.} \beq
\mu_{5}=\frac{8GM}{3\pi}, \eeq with the black hole mass $M$.
Exploiting the Riemann tensors given in (\ref{rabcd}) in Appendix
A, one can readily show that the five-dimensional metric
(\ref{metric5}) is the vacuum solution of the Einstein field
equations for the exterior region of the five-dimensional black
hole. Here, we observe that, in the restriction of $\alpha=\pi/2$,
the five-dimensional Schwarzschild metric (\ref{metric5}) reduces
to the ordinary four-dimensional Schwarzschild submanifold form~\cite{sch} with a
modified gravitational potential originated from the geometric
properties of the five-dimensional Schwarzschild black hole. In
fact, in the weak gravity limit, the gravitational potential in
the five-dimensional theory is proportional to $1/r^{2}$,
different from $1/r$ of the four-dimensional Newtonian theory.
These points will be discussed in more detail in ensuing sections.

Now, we evaluate the horizon radius $r_{H}$ by exploiting the
vanishing lapse function at $r=r_{H}$, to yield \beq
r_{H}=\mu_{5}^{1/2}, \eeq and the surface gravity $k_{H}$ is given
by \beq
k_{H}=\frac{1}{2}\frac{dN^{2}}{dr}|_{r=r_{H}}=\frac{1}{r_{H}}.
\eeq After some tedious algebra, for the five-dimensional
Schwarzschild black hole in the whole region, we obtain the (6+1)
global embedding Minkowski space (GEMS) structure \beq
ds^{2}=-(dz^{0})^{2}+(dz^{1})^{2}+(dz^{2})^{2}+(dz^{3})^{2}+(dz^{4})^{2}+(dz^{5})^{2}+(dz^{6})^{2}
\label{gemsds} \eeq with the coordinate transformations \bea
z^{0}&=&k_{H}^{-1}\left(1-\frac{\mu_{5}}{r^{2}}\right)^{1/2}\sinh k_{H}t,\nn\\
z^{1}&=&k_{H}^{-1}\left(1-\frac{\mu_{5}}{r^{2}}\right)^{1/2}\cosh k_{H}t,\nn\\
z^{2}&=&\int dr~\left(\frac{r_{H}^{2}(r^{3}+r_{H}r^{2}+r_{H}^{2}r+r_{H}^{3})}{r^{4}(r+r_{H})}\right)^{1/2},\nn\\
z^{3}&=&r\sin\alpha\sin\theta\cos\phi,\nn\\
z^{4}&=&r\sin\alpha\sin\theta\sin\phi,\nn\\
z^{5}&=&r\sin\alpha\cos\theta,\nn\\
z^{6}&=&r\cos\alpha. \label{gems} \eea

Exploiting the Christoffel symbols for the five-dimensional
Schwarzschild black hole given in (\ref{chris}) of Appendix A, we
obtain the five-acceleration \beq
a_{d=5}=\frac{r_{H}^{2}}{r^{2}[(r-r_{H})(r+r_{H})]^{1/2}},
\label{ad5} \eeq and the Hawking temperature $T_{d=5}^{H}$ in terms
of the above seven-acceleration $a_{d=7}$ to produce \beq
T_{d=5}^{H}=\frac{a_{d=7}}{2\pi}=\frac{1}{2\pi
r_{H}[1-(r_{H}/r)^{2}]^{1/2}}. 
\label{hawtem1}\eeq For the sake of comparison, we note that 
$T_{d=4}^{H}$ in the standard four-dimensional Schwarzschild black hole is given by~\cite{deser99}
\beq
T_{d=4}^{H}=\frac{1}{4\pi
r_{S}[1-(r_{S}/r)]^{1/2}},
\label{hawtem2} 
\eeq
where $r_{S}=2GM$. The difference between the Hawking temperatures in (\ref{hawtem1}) and (\ref{hawtem2}) 
originates from the fact that the Newtonian force laws in these two cases are different from each other.   

\section{Hydrodynamic properties}
\setcounter{equation}{0}
\renewcommand{\theequation}{\arabic{section}.\arabic{equation}}

In this section, we assume theoretically  at the moment that
the massive particles and photons are moving around the
five-dimensional Schwarzschild black hole to investigate their hydrodynamic
processes. We consider the five
velocity given by $u^{a}=dx^{a}/d\tau$
where one can choose $\tau$ to be the proper time
(affine parameter) for timelike (null) geodesics. From the
equations of motion of a massive particle and/or a photon around the five-dimensional
Schwarzschild black hole, the massive particles initially at rest
and the photons with an initial velocity $u_{\infty}^{r}\approx 1$, respectively.

Now, the fundamental equations of relativistic fluid dynamics can
be obtained from the conservation of particle number and
energy-momentum fluxes.  In order to derive an equation for the
conservation of particle numbers one can use the particle flux
four-vector $nu^{a}$ to yield
\beq
\nabla_{a}(nu^{a})=\frac{1}{\sqrt{-g}}\pa_{a}(\sqrt{-g}~ nu^{a})=0,
\label{fluxcon} \eeq where $n$ is the proper number density of
particles measured in the rest frame of the fluid of massive particles and
photons and $\nabla_{a}$ is the covariant derivative in
the five-dimensional
Schwarzschild curved manifold of interest and the determinant of $g_{ab}$ is
defined by $g={\rm det}~g_{ab}$. For steady state flow of the perfect fluid of the massive
particles and photons, the conservation of
energy-momentum fluxes is similarly described by the Einstein
equation as below \beq
\nabla_{b}T_{a}^{b}=\frac{1}{\sqrt{-g}}\pa_{b}(\sqrt{-g}~
T_{a}^{b})=0, \label{eineqn} \eeq where the stress-energy tensor
$T^{ab}$ for perfect fluid is given by \beq T^{ab}=\rho
u^{a}u^{b}+P(g^{ab}+u^{a}u^{b}), \label{tensor} \eeq with $\rho$
and $P$ being the proper energy densities and pressures of the
massive particles or photons, respectively. In obtaining the first equalities
in (\ref{fluxcon}) and (\ref{eineqn}), we have used the
the Christoffel symbols for the five-dimensional Schwarzschild black hole given in
(\ref{chris}) of Appendix A.

For the steady state flow of the perfect fluid of the
massive particles and photons, the equations
(\ref{fluxcon}) and (\ref{eineqn}) yield \bea 2\pi^{2}
nu^{r}r^{3}\sin^{2}\alpha\sin\theta&=&A_{0},
\label{flow1}\\
(P+\rho)u_{i}u^{r}r^{3}\sin^{2}\alpha\sin\theta&=&A_{i},
\label{flow2}
\eea
where $A_{0}$ is the accretion rate of the massive particles or photons,
and $A_{i}$ ($i=t,\phi$) are the other constants of the motion which can be
evaluated at infinity to yield the ratio $A_{\phi}/A_{t}=u_{\phi}/u_{t}=0$.  Combining the equations (\ref{flow1}) and
(\ref{flow2}), one can derive the relations
\beq \frac{(P+\rho)^{2}}{n^{2}}\left(\kappa\left(1-\frac{\mu_{5}}{r^{2}}\right)+u^{r}u^{r}\right)
=\frac{(P_{\infty}+\rho_{\infty})^{2}}{n_{\infty}^{2}}(\kappa+(1-\kappa)u^{r}_{\infty}u^{r}_{\infty}), \label{prho}
\eeq
where $n_{\infty}$, $P_{\infty}$ and
$\rho_{\infty}$ are the particle number density, pressure and internal energy
density of the fluid of the massive particles or photons at infinity, respectively.
Here we have introduced a new parameter $\kappa$ defined as \beq
\kappa=-g_{ab}u^{a}u^{b}=\left\{\begin{array}{ll}
1 &{\rm for~timelike~geodesics}\\
0 &{\rm for~null~geodesics}.
\end{array}\right.
\label{kappa}
\eeq

Next, using the projection operators in (\ref{eineqn}) one can
obtain the general relativistic equation on the direction
perpendicular to the five-velocity~\cite{wald} \beq
(P+\rho)u^{b}\nabla_{b}u_{a}+(g_{ab}+u_{a}u_{b})\nabla^{b}P=0
\label{euler} \eeq from which, after some algebra, we obtain the
radial component of the above equation for the steady state
axisymmetric accretion of the massive particles and photons on the five-dimensional
Schwarzschild black hole of mass $M$
\beq
\frac{1}{2}\frac{d}{dr}(u^{r}u^{r})+\kappa\frac{\mu_{5}}{r^{3}}
+\frac{1}{P+\rho}\left(u^{r}u^{r}+1-\frac{\mu_{5}}{r^{2}}\right)
\frac{dP}{dr}=0. \label{euler2} \eeq
The equation (\ref{euler2}) is one of the main results, comparable
to the spherically symmetric four-dimensional Schwarzschild black
hole result~\cite{shapiro73} \beq
\frac{1}{2}\frac{d}{dr}(u^{r}u^{r})+\kappa\frac{M}{r^{2}}+\frac{1}{P+\rho}\left(u^{r}u^{r}+1-\frac{2M}{r}\right)
\frac{dP}{dr}=0. \label{eulers} \eeq
The Einstein equation (\ref{eineqn}) can be
easily rewritten in another covariant form \beq
u_{a}\nabla_{b}((P+\rho)u^{b})+(P+\rho
)u^{b}\nabla_{b}u_{a}+\nabla_{a}P=0. \label{eineqn2} \eeq
Multiplying this equation by $u^{a}$ one can project it on the
direction of the four-velocity to obtain \beq
nu^{a}\left(\nabla_{a}\frac{P+\rho}{n}-\frac{1}{n}\nabla_{a}P\right)=0
\label{proj} \eeq where the continuity equation (\ref{fluxcon})
has been used.  The radial component of the equation (\ref{proj})
yields\footnote{This generic result in (\ref{radial}) is 
shared by the Kerr black hole~\cite{kerrfluid}. Here we have included the massless photon case, 
in addition to the massive particle one given in~\cite{shapiro73}.}  
\beq
\kappa\frac{d\rho}{dr}-\kappa\frac{P+\rho}{n}\frac{dn}{dr}+(\kappa-1)\frac{d P}{dr}
=\frac{\Lambda-\Gamma}{u^{r}}.
\label{radial} \eeq Here the energy loss $\Lambda$ and the energy
gain $\Gamma$ are introduced to set the decrease in the entropy of
the inflowing massive and/or massless particles equal to difference 
$\Lambda-\Gamma$.

\section{Effective potential on equatorial plane of
restricted three-brane}
\setcounter{equation}{0}
\renewcommand{\theequation}{\arabic{section}.\arabic{equation}}

In the five-dimensional Schwarzschild metric (\ref{metric5}),
where $t$ and $\phi$ are cyclic, we have two Killing vector fields
$\xi_{i}^{a}~(i=t, \phi)$ satisfying the Killing equations \beq
\pounds_{\xi_{i}}g_{ab}=\na_{a}\xi_{ib}+\na_{b}\xi_{ia}=0. \eeq
Using the above Killing equations and geodesic ones, one can
readily see that $u^{c}\na_{c}(g_{ab}\xi_{i}^{a}u^{b})=0~(i=t,
\phi)$ to produce the constants of motion $\epsilon$ and $l$
corresponding to the Killing vector fields, \bea
g_{ab}\xi_{t}^{a}u^{b}&=&-\left(1-\frac{\mu_{5}}{r^{2}}\right)u^{t}=-\epsilon,\label{gabtaub}\\
g_{ab}\xi_{\phi}^{a}u^{b}&=&r^{2}\sin^{2}\alpha\sin^{2}\theta~u^{\phi}=l,\label{gabphia}
\eea
where $\epsilon$ and $l$ are the conserved energy per unit mass and angular momentum per unit mass, for the massive particles.
For the photons, we note that $\hbar \epsilon$ and $\hbar l$ are the total energy and
the angular momentum of the photons, respectively.

In this section, by taking an ansatz $\alpha=\pi/2$, we consider
the restricted three-brane times ${\mathbf R}$ submanifold defined on the hyper-disk
$S^{2}$ times ${\mathbf R}^{2}$, on which the effective
four-dimensional Schwarzschild metric is described. This four-dimensional Schwarzschild black hole defined 
on the restricted three-brane has a form different from that of the four-metric of the standard 
Schwarzschild back hole, due to the different type of the modified Newtonian inverse square
potential law: \beq
ds_{4}^{2}=-\left(1-\frac{\mu_{5}}{r^{2}}\right)dt^{2}+\left(1-\frac{\mu_{5}}{r^{2}}\right)^{-2}dr^{2}
+r^{2}(d\theta^{2}+\sin^{2}d\phi^{2}). \label{ds42} \eeq
Here we reemphasize that the inverse square potential law holds even on the restricted four-dimensional submanifold. 
To investigate the global geometrical and topological effects of the
five-dimensional Schwarzschild black hole on the effective
potentials on the three-brane, we assume that the test particles
or photons reside on the equatorial plane with $\theta=\pi/2$ of
the restricted three-brane, so that the four-metric (\ref{ds42})
can reduce to the three brane where our effective ordinary
cosmological world is defined, \beq
ds_{3}^{2}=-\left(1-\frac{\mu_{5}}{r^{2}}\right)dt^{2}+\left(1-\frac{\mu_{5}}{r^{2}}\right)^{-2}dr^{2}
+r^{2}d\phi^{2}. \label{ds32} \eeq On this restricted three-brane,
the angular momentum $l$ per unit mass in (\ref{gabtaub}) reduces
to \beq l=r^{2}u^{\phi}. \label{lr2u} \eeq Combining
(\ref{gabtaub}) and (\ref{lr2u}) with (\ref{kappa}), we arrive at
the equation of motion of the massive particles and photons with
the effective potential $V$, as follows \beq
\frac{1}{2}\dot{r}^{2}+V=\frac{1}{2}\epsilon^{2}, \eeq where
$\dot{r}=dr/d\tau$ and \beq
V=\frac{1}{2}\kappa-\kappa\frac{\mu_{5}}{2r^{2}}+\frac{l^{2}}{2r^{2}}-\frac{\mu_{5}l^{2}}{2r^{4}}.
\label{veff} \eeq Here one notes that the second term is the
modified Newtonian term due to the higher dimensional total
manifold geometry, the third is the corresponding centrifugal term
and the last term is the general relativistic term, respectively.

Now, we consider the effective timelike geodesics for the massive
particle constrained in the potential (\ref{veff}) with the
restriction on $l$: $l^{2}>\mu_{5}$. To investigate the
orbit of the particle, we calculate \beq
\frac{dV}{dr}(r=R)=0, \eeq where $R$ is the radius of the unstable orbit \beq
R=\left(\frac{2\mu_{5}}{1-\mu_{5}/l^{2}}\right)^{1/2},
\label{trpm} \eeq 
which is greater than the value of $r_{H}$. Differently from the four-dimensional Schwarzschild 
black hole case where we have the radii of the stable and unstable orbits,
in the five-dimensional Schwarzschild black hole we have only the unstable 
orbit for the massive particle. To see this fact more carefully, we evaluate 
the following quantities
\beq
\frac{d^{2}V}{dr^{2}}(r=R)=-\frac{2(l^{2}-\mu_{5})}{R^{4}},\label{d2vdr2}
\eeq
and
\beq
V(r=R)=\frac{1}{2}+\frac{(l^{2}-\mu_{5})^{2}}{8\mu_{5}l^{2}},\label{d2vdr2}
\eeq
which is greater than $1/2$. The results (\ref{d2vdr2}) and (\ref{d2vdr2}) indicate that 
the V at $r=R$ has the maximum value which is bigger than the asymptotic value $V=1/2$ 
which we can obtain when $r$ approaches the infinity. We thus have a potential barrier with a peak located 
at $r=R$ and the massive particle with $\epsilon>1$ thus possesses an unbound orbit. 
For the case of $l^{2}<\mu_{5}$, there exist no
extrema of $V$ and the massive particle is going down along the
monotonic potential curve up to the black mass. It is interesting
to see that this particle crosses the $V=0$ on the circular ring
at $r=r_{H}$, the event horizon of the five- or four-dimensional
Schwarzschild black hole. After some algebra in our case at hand, we can readily obtain 
the impact parameter $b$ as follows,
\beq
b^{2}=\frac{\epsilon^{2}-1}{\epsilon^{2}}\cdot\frac{R^{4}}{\mu_{5}}.
\eeq

Next, we consider the null geodesics of photon whose potential is given by inserting 
$\kappa=0$ into (\ref{veff}) as follows
\beq
V=\frac{l^{2}}{2r^{2}}-\frac{\mu_{5}l^{2}}{2r^{4}}.
\label{veff2} 
\eeq 
The orbit of the photon can be now obtainable by considering the following extrema condition
\beq
\frac{dV}{dr}(r=R_{0})=0, \eeq where $R_{0}$ is the radius of the unstable orbit \beq
R_{0}=\left(2\mu_{5}\right)^{1/2},
\label{trpm} \eeq 
which is also greater than the value of $r_{H}$. 
Similar to the four-dimensional Schwarzschild 
black hole, we have the unstable photon orbit radius $r=R_{0}$ in the five-dimensional 
Schwarzschild black hole. Moreover, we can readily check that $d^{2}V/dr^{2}(r=R_{0})=-6\mu_{5}l^{2}/R_{0}^{6}$ 
and $V(r=R_{0})=l^{2}/8\mu_{5}$ indicating the peak of the potential (\ref{veff2}) at $r=R_{0}$.
Here one notes that on the black hole horizon at $r=r_{H}$, the potential $V(r=r_{H})$ vanishes in both cases of the 
massive particle and photon. The impact parameter $b$ for the photon is now given by,
\beq
b^{2}=\frac{R_{0}^{4}}{R_{0}^{2}-\mu_{5}}.
\eeq

\section{Conclusions}
\setcounter{equation}{0}
\renewcommand{\theequation}{\arabic{section}.\arabic{equation}}

We have introduced the five-dimensional Schwarzschild black hole metric to obtain the (6+1) dimensional GEMS 
structure and the five-acceleration of this higher dimensional black hole. Here one notes that the 
four-dimensional Schwarzschild black hole has the (5+1) dimensional GEMS structure~\cite{deser97,deser99}. 
We have next evaluated its thermodynamic physical quantity such as the Hawking temperature on this five-dimensional manifold. 
Here we have noticed that the five-dimensional Hawking temperature has the form different from the four-dimensional one due to the 
fact that the Newtonian force laws in these two cases are different from each other.

Next, we have investigated the hydrodynamic aspects of the five-dimensional Schwarzschild black hole, by assuming that 
the massive particles and photons are moving around the black hole. We have obtained the radial component equation for 
the the steady state axisymmetric accretion of the massive particles and photons on the five-dimensional
Schwarzschild black hole, and this radial equation has been shown to be different from that of the standard four-dimensional 
Schwarzschild black hole, only because of the differences in their force laws. We then have found the radial 
component of the Einstein equation associated with the entropies of the massive particles and photons. Remarkably, this 
equation has shown to be the same form as the four-dimensional black hole cases.

Finally, we have assumed that the test particles or photons reside on the equatorial plane with $\theta=\pi/2$ of
the restricted three-brane with $\alpha=\pi/2$, namely, we could have the three-brane on which the effective ordinary
cosmological world is defined. On this restricted brane, we have studied the behaviors 
of the particles or photons. 

\appendix
\section{Geometry of five-dimensional Schwarzschild black hole}
\setcounter{equation}{0}
\renewcommand{\theequation}{A.\arabic{equation}}

Using the five-dimensional Schwarzschild black metric
(\ref{metric5}), we list the nonvanishing Christoffel symbols
which have been exploited in the evaluation of the five-acceleration
$a_{d=5}$ in (\ref{ad5}) \beq
\begin{array}{ll}
\Gamma_{tr}^{~t}=-\Gamma_{rr}^{~r}=\frac{\mu_{5}}{r(r^{2}-\mu_{5})},
&\Gamma_{tt}^{~r}=\frac{\mu_{5}(r^{2}-\mu_{5})}{r^{5}},\\
\Gamma_{\alpha\alpha}^{~r}=-\frac{r^{2}-\mu_{5}}{r},
&\Gamma_{\theta\theta}^{~r}=-\frac{(r^{2}-\mu_{5})\sin^{2}\alpha}{r},\\
\Gamma_{\phi\phi}^{~r}=-\frac{(r^{2}-\mu_{5})\sin^{2}\alpha\sin^{2}\theta}{r},
&\Gamma_{r\alpha}^{~\alpha}=\Gamma_{r\theta}^{~\theta}=\Gamma_{r\phi}^{~\phi}
=\frac{1}{r},\\
\Gamma_{\theta\theta}^{~\alpha}=-\sin\alpha\cos\alpha,
&\Gamma_{\phi\phi}^{~\alpha}=-\sin\alpha\cos\alpha\sin^{2}\theta,\\
\Gamma_{\alpha\theta}^{~\theta}=\Gamma_{\alpha\phi}^{~\phi}=\cot\alpha,
&\Gamma_{\phi\phi}^{~\theta}=-\sin\theta\cos\theta,\\
\Gamma_{\theta\phi}^{~\phi}=\cot\theta. &\\
\label{chris} 
\end{array}
\eeq For
the sake of completeness, we calculate the nonzero Riemann tensors
\beq
\begin{array}{ll}
R_{trtr}=-\frac{3\mu_{5}}{r^{4}}, &R_{t\alpha t\alpha}=\frac{\mu_{5}(r^{2}-\mu_{5})}{r^{4}},\\
R_{t\theta t\theta}=\frac{\mu_{5}(r^{2}-\mu_{5})\sin^{2}\alpha}{r^{4}}, 
&R_{t\phi t\phi}=\frac{\mu_{5}(r^{2}-\mu_{5})\sin^{2}\alpha\sin^{2}\theta}{r^{4}},\\
R_{r\alpha r\alpha}=-\frac{\mu_{5}}{r^{2}-\mu_{5}}, &R_{r\theta r\theta}=-\frac{\mu_{5}\sin^{2}\alpha}{r^{2}-\mu_{5}},\\
R_{r\phi r\phi}=-\frac{\mu_{5}\sin^{2}\alpha\sin^{2}\theta}{r^{2}-\mu_{5}}, 
&R_{\alpha\theta\alpha\theta}=\mu_{5}\sin^{2}\alpha,\\
R_{\alpha\phi\alpha\phi}=\mu_{5}\sin^{2}\alpha\sin^{2}\theta, 
&R_{\theta\phi\theta\phi}=\mu_{5}\sin^{4}\alpha\sin^{2}\theta,
\label{rabcd} 
\end{array}
\eeq which yield the vanishing Ricci tensors and the
corresponding vanishing scalar curvature. These tensors and the
curvature have been used in Section II.

\end{document}